\documentclass[aps,prl,preprintnumbers,amsmath,amssymb,latexsym,array,enumerate,letter,twocolumn,superscriptaddress]{revtex4}
\usepackage{amssymb}
\usepackage{amsmath}
\usepackage{epsfig}
\usepackage{hyperref}
\usepackage{breakurl}
\usepackage{xcolor}

\makeatletter
\def\simgt{\mathrel{\lower2.5pt\vbox{\lineskip=0pt\baselineskip=0pt
           \hbox{$>$}\hbox{$\sim$}}}}
\def\simlt{\mathrel{\lower2.5pt\vbox{\lineskip=0pt\baselineskip=0pt
           \hbox{$<$}\hbox{$\sim$}}}}
\makeatother

\newcommand{\be}{\begin{equation}}
\newcommand{\ee}{\end{equation}}
\newcommand{\bea}{\begin{eqnarray}}
\newcommand{\eea}{\end{eqnarray}}
\newcommand{\beq}{\begin{eqnarray}}
\newcommand{\eeq}{\end{eqnarray}}

\usepackage{titlesec}
\titleformat{\section}[runin]{\normalfont \bfseries}{\thesection}{1em}{}

\def\lsim{\mathrel{\rlap{\lower4pt\hbox{\hskip1pt$\sim$}}
     \raise1pt\hbox{$<$}}}         
\def\gsim{\mathrel{\rlap{\lower4pt\hbox{\hskip1pt$\sim$}}
     \raise1pt\hbox{$>$}}}         

\begin{document}


\title{Boosted Dark Matter Interpretation of the XENON1T Excess}

\author{Bartosz Fornal}
\affiliation{\mbox{Department of Physics and Astronomy,
University of Utah, Salt Lake City, UT 84112, USA}}

\author{Pearl Sandick}
\affiliation{\mbox{Department of Physics and Astronomy,
University of Utah, Salt Lake City, UT 84112, USA}}

\author{Jing Shu}
\affiliation{CAS Key Laboratory of Theoretical Physics, Institute of Theoretical Physics,
Chinese Academy of Sciences, Beijing 100190, China.}
\affiliation{School of Physical Sciences, University of Chinese Academy of Sciences, Beijing 100049, P. R. China.}
\affiliation{CAS Center for Excellence in Particle Physics, Beijing 100049, China}
\affiliation{Center for High Energy Physics, Peking University, Beijing 100871, China}
\affiliation{School of Fundamental Physics and Mathematical Sciences, Hangzhou Institute for Advanced Study, University of Chinese Academy of Sciences, Hangzhou 310024, China}
\affiliation{International Centre for Theoretical Physics Asia-Pacific, Beijing/Hangzhou, China}

\author{Meng Su}
\affiliation{Department of Physics, The University of Hong Kong, Hong Kong SAR, China}
\affiliation{Laboratory for Space Research, The University of Hong Kong, Hong Kong SAR, China\vspace{2mm}}

\author{Yue Zhao\vspace{2mm}}
\affiliation{\mbox{Department of Physics and Astronomy,
University of Utah, Salt Lake City, UT 84112, USA}}

\begin{abstract}
We propose boosted dark matter (BDM) as a possible explanation for the excess of keV electron recoil events observed by XENON1T. BDM particles have velocities much larger than those typical of virialized dark matter, and, as such, BDM-electron scattering can naturally produce keV electron recoils. We show that the required BDM-electron scattering cross sections can be easily realized in a simple model with a heavy vector mediator. Though these cross sections are too large for BDM to escape from the Sun, the BDM flux can originate from the Galactic Center or from halo dark matter annihilations. Furthermore, a daily modulation of the BDM signal will be present, which could not only be used to differentiate it from various backgrounds, but would also provide important directional information for the BDM flux. \vspace{2mm}

\end{abstract}

\maketitle

\section{Introduction}

\hspace{-2mm}The XENON1T experiment has recently reported an excess in their low energy electron recoil data, appearing between $2-3$ keV~\cite{Aprile:2020tmw}. Three possible\break explanations are proposed in~\cite{Aprile:2020tmw} for this excess. The first possible explanation is beta decays of tritium, which currently can neither be confirmed nor ruled out due to our lack of knowledge of the tritium concentration. 
The other two possible explanations pursued in~\cite{Aprile:2020tmw} are solar axions and anomalous neutrino interactions \cite{Bell:2005kz,Bell:2006wi}.  However, the preferred values of couplings in the latter two cases have already been ruled out by existing astrophysical constraints, particularly from stellar cooling~\cite{Raffelt, Giannotti:2017hny, Corsico:2014mpa, 2019arXiv191010568A}. 
Furthermore, it is well known that the observed excess in electron recoils cannot be due to typical dark matter (DM) particles scattering on electrons, since a DM particle moving at virial velocity, i.e. ${\mathcal{O}}(10^{-3}) \,c$, will result in an energy deposition that is much smaller than the keV-scale excess. Therefore it is interesting and important to explore other possible explanations for the observed excess. 

Many theoretical models predict the existence of boosted dark matter (BDM) in our Universe. With its typical velocities much larger than the virial velocity, BDM models are therefore generically capable of producing an excess in electron recoils such as the one observed.  For example, the DM-induced nucleon decay process studied in~\cite{Huang:2013xfa} produces BDM in its final state, while semi-annihilation~\cite{DEramo:2010keq} and multi-component~\cite{Agashe:2014yua} DM models also yield BDM fluxes from the Sun or the Galactic Center (GC). BDM flux searches have been proposed for large volume neutrino experiments, and interesting parameter space has been covered by existing experiments, such as Super Kamiokande \cite{Kachulis:2017nci}, ProtoDUNE \cite{Chatterjee:2018mej} and IceCube \cite{Bhattacharya:2016tma,Kopp:2015bfa}, or will be covered by future experiments such as DUNE \cite{Berger:2019ttc,Arguelles:2019xgp,Abi:2020evt}; for other related work on BDM see \cite{Kamada:2017gfc,Kamada:2018hte,McKeen:2018pbb,Kamada:2019wjo}.


In this study, we propose that the keV excess in electron recoils observed by XENON1T could be due to BDM scattering on electrons. If $m_{\rm BDM}\gg m_e$, a typical energy deposition from BDM-electron scattering of $\sim$ few keV implies the velocity of the BDM must be ${\mathcal{O}}(0.1) \,c$, thus only mildly boosted. Here we present two example models, and consider the flux of BDM particles from the Milky Way halo and the Sun. We find that including BDM-electron recoils can significantly improve the fit to the data relative to the background only hypothesis, and that the required scattering cross section can be naturally explained by models with a vector mediator. 
Finally, we highlight that our BDM signals enjoy a daily modulation. Such a unique feature can be used to distinguish the signal proposed here from various backgrounds, as well as many other new physics interpretations of this excess \cite{Smirnov:2020zwf,Takahashi:2020bpq,Kannike:2020agf,Chen:2020gcl}.  Furthermore, the phase and the amplitude of the modulation can be used to extract crucial information about the BDM, e.g. its flux direction, providing essential guidance for a future experimental analysis.





\section{Models for Boosted Dark Matter}

DM particles moving at virial velocity are not capable of depositing energy as large as a few keV when scattering on electrons. However, in the BDM scenario, some fraction of DM particles are boosted such that their velocities are much larger than those typical of virialized DM.  Indeed, BDM scattering on electrons would therefore result in higher energy recoil signals.  The existence of a BDM flux is a generic prediction in many well-motivated DM models.  In this section, we present two example models in which a BDM flux appears naturally.

The first example model is the semi-annihilation DM model \cite{DEramo:2010keq}. In this case, DM $\chi$ carries a $Z_3$ symmetry, and the BDM flux is produced through the following annihilation process: 
\begin{eqnarray}\label{semi}
\chi+\chi\to\bar\chi+X.
\end{eqnarray}
Here $X$ represents a particle that does not carry $Z_3$ charge, which can be a SM particle or can eventually decay to SM particles. The boost factor of the $\bar\chi$ in the final state is $\gamma_\chi={(5m_\chi^2-m_X^2)}/{4m_\chi^2}$.
In the limit of $m_\chi\gg m_X$, the boost factor reaches its maximum value of 1.25. 

The second BDM model we study here is the two-component DM model (see, e.g.~\cite{Boehm:2003ha} and subsequent work).  In this case, two particles $\psi_A$ and $\psi_B$ are both stable. We assume $\psi_A$ is heavier and is the dominant component of DM. The annihilation of $\psi_A$ particles produces boosted $\psi_B$ particles in the final state,
\begin{eqnarray}\label{two-comp}
\psi_A+\bar\psi_A\to \psi_B+\bar\psi_B.
\end{eqnarray}
The boost factor of the $\psi_B$ particles is simply the mass ratio, i.e. $\gamma_B=m_A/m_B$.

\section{Boosted Dark Matter Sources}

There are two promising sources to generate the BDM flux: annihilation in the GC/halo, as well as capture and annihilation in the Sun. Here we summarize the expected flux from each source.

Assuming the DM follows an NFW profile~\cite{Navarro:1995iw}, the BDM flux from the full sky can be written as~\cite{Agashe:2014yua}
\begin{eqnarray}\label{two-comp}
\Phi^{\rm BDM}_{\rm gal} &=& 1.6\times 10^{-6} \, {\rm cm}^{-2}{\rm s}^{-1} \nonumber \\
&\,&\times\bigg(\frac{\langle\sigma_{\rm ann} v\rangle}{5\times 10^{-26}\,{\rm cm}^3{\rm s}^{-1}}\bigg)\bigg(\frac{10 \, {\rm GeV}}{m_{{\rm DM}}}\bigg)^2,
\end{eqnarray}
where $\langle\sigma_{ann} v\rangle$ is the total thermally-averaged DM annihilation cross section at present time. We note that DM can also be produced in a non-thermal manner, in which case the DM annihilation cross section can be larger, leading to a larger BDM flux from the GC.  
Though the DM density peaks towards the GC, since XENON1T cannot distinguish the direction of the incoming DM particle, all sky directions should be included.  

A second source for the BDM flux is the Sun.  If DM particles scatter on nuclei and are captured by the Sun, DM can accumulate in the Sun's core over time. The solar capture rate can be approximated by~\cite{Berger:2014sqa}
\begin{eqnarray}\label{capture}
C(m_{\rm DM},\sigma_{\rm nucl}) &\simeq&  2\times 10^{22}\,{\rm s}^{-1} \nonumber \\
&\,& \times\bigg(\frac{\sigma_{\rm nucl}}{10^{-42}\,{\rm cm}^2}\bigg)\bigg(\frac{10 \, {\rm GeV}}{m_{{\rm DM}}}\bigg)^2.
\end{eqnarray}
For simplicity, this approximation assumes that the DM-nucleon scattering cross section does not depend on the relative velocity at leading order.  If the leading order cross section has a $v^2$ dependence, the DM capture rate can be enhanced by a factor of $\sim25$.  
We also note that our choice of benchmark value for the DM-nucleon scattering cross section of $\sigma_{\rm nucl}=10^{-42} \, {\rm cm}^2$ corresponds to the bound on the spin-dependent scattering cross section obtained by DM direct detection experiments, see, e.g. \cite{Aalbers:2016jon}.

For typical choices of DM scattering and annihilation cross sections, the Sun will reach a capture-annihilation equilibrium \footnote{For a detailed discussion, please see~\cite{Berger:2014sqa}.}. In this case, the DM annihilation cross section becomes irrelevant, and the BDM flux is fully determined by the DM capture rate, which is characterized by the DM-nucleon scattering cross section, $\sigma_{\rm nucl}$.  Thus, the BDM flux can be written as
\begin{eqnarray}\label{capture}
\Phi^{\rm BDM}_{\rm Sun} &=& \frac{C(m_{\rm DM},\sigma_{\rm nucl})}{4\pi \, {\rm AU}^2}\\
&=&7.2\times 10^{-6} \,{\rm cm}^{-2} {\rm s}^{-1}\bigg(\frac{\sigma_{\rm nucl}}{10^{-42} \, {\rm cm}^2}\bigg)\bigg(\frac{10 \, {\rm GeV}}{m_{\rm DM}}\bigg)^2\nonumber,
\end{eqnarray}
where AU is an astronomical unit.  We note that there is an important subtlety regarding the BDM flux from the Sun, to which we will return shortly.

\section{Signal Rate}

For a given BDM flux, one can estimate the total number of signal events as
\begin{eqnarray}\label{event}
N_{\rm sig} &=& Z'\, n_{\rm Xe} \,V\, T\, \sigma_{\rm elec}\, \Phi^{\rm BDM} \nonumber\\
&=&Z'\,  \frac{M_{\rm det}T}{m_{\rm Xe}}\times \sigma_{\rm elec} \times \Phi^{\rm BDM} \ .
\end{eqnarray}
Here $n_{\rm Xe}$ is the number density of xenon atoms in the detector, $M_{\rm det}$ and $V_{\rm det}$ are the fiducial mass and volume of the detector, $T$ is the total operation time, and $\sigma_{\rm elec}$ is the BDM-electron scattering cross section. The exposure $M_{\rm det}T$ for the XENON1T data presented in~\cite{Aprile:2020tmw} is 0.65 tonne-years. $Z'$ is the effective number of electrons in xenon that undergo recoils. In particular: electrons on shells $K$ and $L$ have binding energies $E_b>4.5 \ {\rm keV}$ and cannot be knocked out by DM with kinetic energies relevant for the excess; for shell $M$ the binding energies  fall within the range $0.68 -1.15 \ {\rm keV}$, leading to contributions at the lower  part  of the spectrum; finally, the electrons on  shells $N$ and $O$ have $E_b< 0.22 \ {\rm keV}$ and contribute almost like free particles. This leads to  $Z' \sim 40$.

In order to explain the excess observed by XENON1T, the number of the signal events needs to be ${\mathcal{O}}(100)$. This translates to a BDM-electron scattering cross section of
\begin{eqnarray}\label{sigma-elec}
 \sigma_{\rm elec} = 3\times 10^{-29} \,{\rm cm}^{2}\bigg(\frac{10^{-6}\,{\rm cm}^{-2}{\rm s}^{-1}}{\Phi^{\rm BDM}}\bigg)\bigg(\frac{N_{\rm sig}}{100}\bigg).
\end{eqnarray}
This provides a rough prediction for the BDM-electron scattering cross section.  

Now, let us examine whether it is reasonable to expect $\sigma_{\rm elec} $ as large as $\mathcal{O}(10^{-29}-10^{-28}) \,{\rm cm}^{2}$.  If a BDM particle scatters with an electron through a vector mediator whose mass is much larger than the typical momentum transfer, the scattering cross section can be written as~\cite{Joglekar:2019vzy}
\begin{eqnarray}\label{sigma-med}
 \sigma_{\rm elec}= \frac{g_{\rm BDM}^2 g_e^2 m_e^2}{\pi \,m_{\rm med}^4}\,,
\end{eqnarray}
where $g_{\rm BDM}$ ($g_e$) is the coupling between the mediator and BDM (electron). The constraints on such a mediator with invisible decay can be found in~\cite{Essig:2013lka}. As a benchmark, consider $g_{\rm BDM}=1.1$,  $g_{e}=10^{-5}$, $m_{\rm BDM}=10$ GeV, and $m_{\rm med}=0.1$ MeV, which results in $\sigma_{\rm elec} =4\times 10^{-29} \,{\rm cm}^2$.  Thus, for reasonable parameter values, a cross section as large as $\sigma_{\rm elec} = \mathcal{O}(10^{-29}-10^{-28}) \,{\rm cm}^{2}$ is easily obtained~\footnote{In semi-annihilation models, there is only one species of DM particle, and the DM capture may be dominantly caused by DM-electron scattering. This will modify the capture rate estimation in Eq. (\ref{capture}). In addition, when the BDM particle is the same as DM particle, such a strong interaction with electron has already been ruled out by DM direct detection experiments, e.g. \cite{Essig:2012yx,Aprile:2019xxb,Barak:2020fql}, unless one introduces non-trivial velocity dependence in DM-electron cross section. Thus semi-annihilation models are disfavored in explaining the XENON1T excess. Lastly, the tension between the light mediator and cosmology can be avoided by a late time phase transition in the dark sector \cite{Chacko:2004cz,Davoudiasl:2017jke,Zhao:2017wmo}.}.

One important question for mildly boosted DM with large $\sigma_{\rm elec}$ is whether the BDM can penetrate the Sun after its production near the core. For a solar core density of $150 \,{\rm g}/{\rm cm}^3$, the free streaming length in the Sun is 
\begin{eqnarray}\label{freestream}
L_{fs,S}\simeq 1 \,{\rm m} \times \bigg(\frac{10^{-28} \, {\rm cm}^2}{\sigma_{\rm elec}}\bigg),
\end{eqnarray}
and for each scattering the momentum transfer is $\mathcal{O}(10-100)$ keV.  With a solar core radius of $1.4\times 10^5$ km, it is  unlikely that the BDM produced near the center of the Sun will escape.  Thus, if the XENON1T excess is explained by BDM, the flux is likely to be produced in the GC, as there would be no appreciable flux from the Sun for the required BDM-electron scattering cross section. We note that simple variations of BDM models considered here can avoid the difficulty of escaping the Sun, which we will comment on later.

Similarly, let us also calculate how far the BDM can propagate in the Earth. Taking the average Earth density as $5.5 \,{\rm g/cm^3}$, the free streaming length in the Earth is 
\begin{eqnarray}\label{freestream-earth}
L_{fs,E}\simeq 60 \,{\rm m} \times \bigg(\frac{10^{-28} \, {\rm cm}^2}{\sigma_{\rm elec}}\bigg).
\end{eqnarray}
The XENON1T experiment operates underground~\footnote{The average rock density above XENON1T is somewhat smaller than the average Earth density quoted here.  The average used here is sufficient for the following order-of-magnitude estimates.} at a depth of about 1600 m. As discussed above, BDM has a velocity of $\mathcal{O}(0.1)\,c$. If every BDM-electron collision reduces the BDM momentum by $\mathcal{O}(10-100)$ keV, a BDM particle whose mass is smaller than $\mathcal{O}(0.01-0.1)$ GeV may not be able to reach the detector, assuming the benchmark BDM-electron scattering cross section at $10^{-28} \,{\rm cm}^2$. Furthermore, if BDM propagates for a distance comparable to the Earth's radius of 6000 km, its momentum may be reduced by $\mathcal{O}(1-10)$ GeV. If the BDM mass is below $\mathcal{O}(10-100)$ GeV, there is a high chance that it cannot fully penetrate the Earth. This leads to a daily modulation of the DM signal if the BDM flux is dominantly from the GC~\footnote{Such Earth shielding effect is under investigation in \cite{shadow}.}. This feature may provide an important handle to reduce various backgrounds and can potentially be used as a smoking gun signature for BDM discovery. More details about daily modulation will be provided in a later discussion.

\section{Energy Deposition Distribution}

\begin{figure}[t!]
\includegraphics[trim={1.5cm 0.4cm 1.5cm 0.2},clip,width=8.8cm]{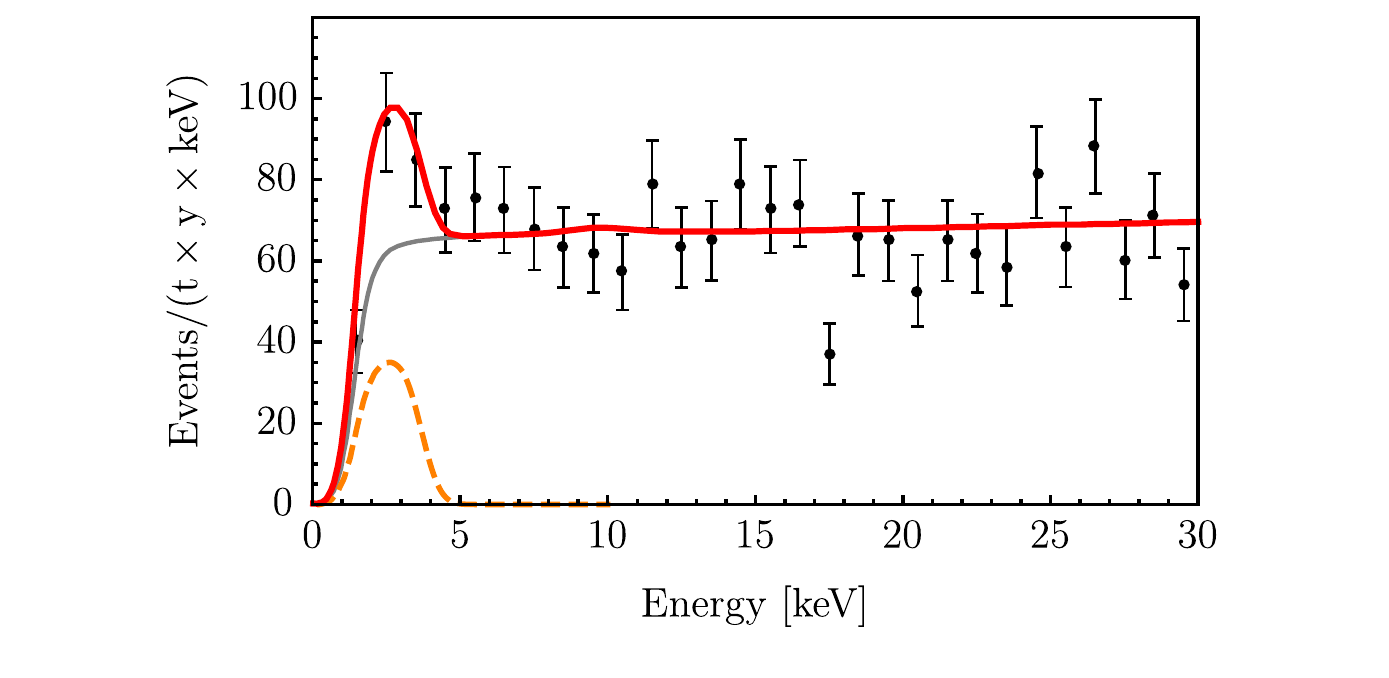}
\caption{The energy spectrum of electrons for a benchmark with $m_{\rm BDM}=10 \ {\rm GeV}$, 
$v_{\rm BDM}=0.06\,c$\,, $\sigma_{\rm elec} = 4\times 10^{-29} \,{\rm cm}^2$ and BDM flux $\Phi^{\rm BDM} \!=\! 10^{-6}\,{\rm cm}^{-2}{\rm s}^{-1}$. 
The dashed orange line represents the contribution from BDM-electron scattering after including the energy resolution and detection efficiency. The red line shows the total electron energy distribution at XENON1T.\\}\label{DM}
\end{figure}  

Finally, we study the electron recoil energy distribution at XENON1T resulting from BSM-electron scattering in the benchmark scenario discussed above. 
The energy deposition required to fit the excess is $\sim$ few keV. 
In the limit $m_{\rm BDM}\gg m_e$, which is applicable in our case, the electron recoil energy is $E_e \leq 2 m_e v_{\rm BDM}^2$. The corresponding differential cross section is approximately a flat function, which can be written as
\bea
\frac{d\sigma_{\rm elec}}{d E_e} = {\rm const}\times \Theta(2 m_e v_{\rm BDM}^2-E_e) \ ,
\eea
where $\Theta$ is the Heaviside step function.

To derive the expected signal at XENON1T, we convolute this differential cross section with a Gaussian of width $\Delta = 0.5 \ {\rm keV}$, corresponding to the detector energy resolution in the $\sim \rm keV$  region \cite{Aprile:2020yad},  and weight the cross section by the detector efficiency \cite{Aprile:2020tmw}. 

In Fig.\,\ref{DM}, we demonstrate how BDM-electron scattering can improve the fit to data. Here we choose as a benchmark: $m_{\rm BDM}=10 \ {\rm GeV}$, $v_{\rm BDM}=0.06\,c$ and $\sigma_{\rm elec} = 4\times 10^{-29} \,{\rm cm}^2$. The BDM-electron scattering produces a bump, denoted by the orange dashed line, in the electron recoiling energy spectrum. The total electron spectrum at XENON1T is given by the red line. For the benchmark model under consideration, adding contributions from BDM-electron scattering can provide an excellent fit to data.

\section{Daily Modulation}

DM signals in direct detection experiments may present non-trivial time dependence, as first studied in~\cite{Drukier:1986tm}, where it was proposed that an annual modulation in scattering events could arise due to the Earth's motion around the Sun.
A daily modulation in DM-nucleus scattering due to the Earth's shielding of a terrestrial detector from the DM wind was explored in~\cite{Collar:1992qc}, and similar cases have been further studied in e.g.~\cite{Hasenbalg:1997hs, Kouvaris:2014lpa, Kavanagh:2016pyr,Emken:2019tni}.

Given the sizable DM-electron scattering cross section necessary to explain the XENON1T excess, a daily modulation of the signal, due to the anisotropy of the BDM flux and the dynamics of Earth's shielding, is a natural prediction. Indeed, a similar daily modulation due to Earth's shielding of a flux of cosmic ray boosted DM from the GC was recently studied in~\cite{Ge:2020yuf}.  

In this section, we demonstrate some features of the expected daily modulation that would allow extraction of information about the BDM model from the XENON1T signal.
We note that the estimates presented here are crude, but serve to illustrate the capability of learning about BDM from a direct detection signal.  Specifically, we consider the Earth to be a sphere of constant density and assume that the XENON1T detector is buried 1.6 km beneath the surface at the location of the Gran Sasso National Lab.   
A more extensive analysis is beyond the scope of this study, but will ultimately be critical for investigating a BDM origin of the XENON1T excess.

\begin{figure}[t!]
\includegraphics[width=7.8cm]{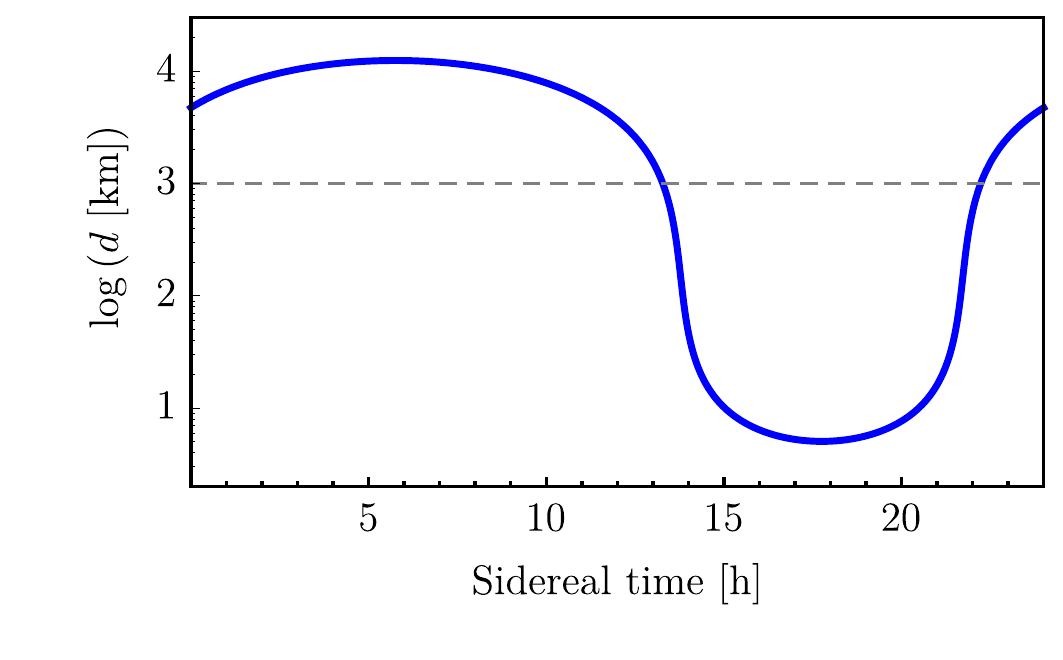}
\includegraphics[width=7.8cm]{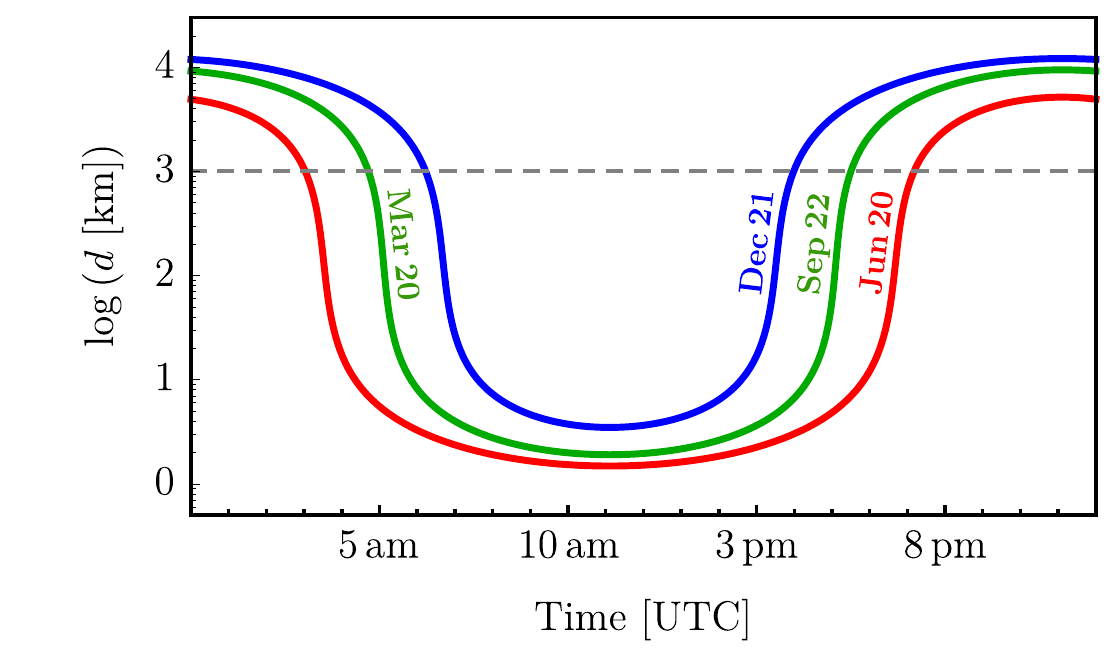}
\caption{{\it Upper:} The depth that a BDM particle from the GC must penetrate to reach the XENON1T detector as a function of the sidereal time.    {\it Lower:} The depth that a BDM particle from the Sun must penetrate to reach the XENON1T detector as a function of the universal time for four dates throughout the year.  
The gray dashed line in each panel indicates a depth of 1000 km, as in Eq.~(\ref{depth}).}\label{daily}
\end{figure} 

First, we consider a BDM flux from the GC. The XENON1T lab rotates around the Earth's axis once each sidereal day, and its position can be parameterized by the sidereal time (ST). In the upper panel of Fig.~\ref{daily}, we show the depth of the Earth, $d$, that a BDM particle must penetrate in order to reach the XENON1T detector, as a function of ST. It is straightforward to convert ST to the local time in Gran Sasso at any given time of the year; the position of the minimum depth shifts by $\sim$ two hours per month  in terms of the local time. The depth $d$ reaches its minimum value when the detector is oriented with the least Earth shielding in the direction of the source, which in this case is the GC. Therefore, the directional information is encoded in the position of the signal peak, i.e. the phase.  Even for a detector such as XENON1T, with no design capability to distinguish directionality, it may be possible to identify the source of a BDM flux.


It may happen, however, that the dominant component of the BDM flux does not come from the GC, but rather from another direction in the sky.
As discussed above, BDM particles arising in the simplest scenarios are unlikely to escape the Sun due to their interactions with electrons, but it is not difficult to construct a BDM model for which a substantial flux from the Sun is expected.
For example, in a multi-component DM model, consider the possibility that $\psi_A$'s are captured by the Sun and annihilate with each other through the decay chain $$\psi_A+\bar \psi_A\to \phi +\phi^* \to 2\psi_B+2\bar \psi_B,$$ where $\phi$ is an intermediate particle whose decay length is comparable to the solar radius. The $\psi_B$'s arising from $\phi$ decay are therefore produced outside the Sun. In this case, the BDM particles would point back to the Sun and could still have a large scattering cross section with electrons.  

In the lower panel of Fig.~\ref{daily}, we show the depth of the Earth $d$ that the BDM must penetrate, as a function of the Universal Coordinated Time (UTC), assuming the Sun is the source of the BDM flux. While the position of the signal peak remains constant, the shape of the modulation varies throughout the year. In the winter, when the days are shortest in the northern hemisphere, XENON1T experiences its shortest daily exposures with minimal shielding, as shown by the relatively short daytime dip in the blue curve in Fig.~\ref{daily}. In the summer, on the other hand, the daily exposure with low shielding is the longest, as demonstrated by the red curve in Fig.~\ref{daily}.  We note that these general conclusions hold independently of the details of the XENON1T position or the Earth density modeling.

Finally, it is instructive to estimate the depth at which most BDM particles are stopped. We assume that each BDM-electron scattering reduces the BDM kinetic energy by $\sim$ 3 keV. This implies that the BDM signal approaches zero when 
\begin{eqnarray}\label{depth}
d_e \gtrsim 1000\ {\rm km} \bigg(\frac{10^{-28}\,{\rm cm^2}}{\sigma_{\rm elec}}\bigg) \bigg(\frac{m_{\rm BDM}}{10\ {\rm GeV}}\bigg)\bigg(\frac{v_{\rm BDM}}{0.1\ c}\bigg)^2. \ \ 
\end{eqnarray}
As an illustration, consider the survival probability $P$ as a function of the penetration depth $d$ to be modeled by a Heaviside step function:~$P(d) = \Theta(d_e-d)$.
The daily modulation of the signal is determined by the survival probability of the  incoming BDM particles.  To make a realistic prediction for the signal, care must be taken when modeling the position and overburden of the detector.
Furthermore, for the BDM flux from the Sun, one must also account for the seasonal variation of $d(t)$, as described above. 
A detailed analysis, including a realistic simulation of the survival probability function for BDM-electron interactions is left for future study.

Again, we emphasize that even for a DM direct detection experiment such as XENON1T which is not designed to detect the directionality of scattering events~\cite{Mayet:2016zxu}, the phase of the daily modulation could be used to extract directional information for the BDM flux.  The GC and the Sun may be the most likely sources for a BDM flux, but it is also conceivable that a BDM flux comes from a nearby concentration of DM, such as a DM sub-halo or a mini-cluster.  Indeed, the phase of the daily modulation could in principle take any value and could be used to identify the flux direction/source. 


\section{Conclusion}
In this study, we explain the XENON1T excess reported in \cite{Aprile:2020tmw} as a signal of BDM-electron scattering. Such an interaction can be naturally introduced through a vector mediator whose mass is much larger than the typical momentum transfer in the scattering. With reasonable choices of parameters, cross sections of the magnitude necessary to explain the excess can be easily obtained and significantly improve the fit to the XENON1T data. We consider two possible sources for the BDM flux: the GC/halo and the Sun. We find that the required BDM-electron scattering cross section is large enough that the BDM particles studied here would not escape from the Sun. Thus, if XENON1T is indeed observing BDM, either it must come from the GC or the Milky Way halo, or the BDM model must be more complicated than the minimal scenario we highlight here. 




Finally, the predicted daily modulation of the signal opens a brand new strategy to investigate possible sources for the excess. The daily modulation will not only improve the signal significance  and help discriminate against backgrounds such as beta decays of tritium, but it also provides a unique opportunity to determine the origin of the BDM flux.
The maximum of the BDM flux occurs when the detector is oriented with the least Earth shielding in the direction of the source.
Therefore, the phase of this daily modulation could potentially resolve the direction of the BDM flux, presenting a novel opportunity to discover the true nature of dark matter.
\vspace{4mm}

{\bf Note added.} -- During the preparation of this work, Ref.\,\cite{Kannike:2020agf} appeared and also interpreted the XENON1T excess using fast moving DM particles. The authors considered the sources as fast moving DM sub-haloes, solar/earth captured semi-annihilating DM, or nearby axion stars. Compared to  \cite{Kannike:2020agf}, we systematically examined the BDM scenarios and carefully studied the BDM flux from the GC and the Sun, pointing out several interesting features and subtleties. We note that several of the benchmarks provided in \cite{Kannike:2020agf} are unlikely to be viable. 
\vspace{-2mm}

\section{Acknowledgments}
We thank Jianglai Liu and Qiang Yuan for helpful discussions. The work of P. S. is supported by NSF grant PHY-1720282. B.F. and Y.Z. are supported by the U.S. Department of Energy under Award No.\,${\rm DE}$-${\rm SC0009959}$. J.S. is supported by the National Natural Science Foundation of China (NSFC) under grant No.11947302, No.11690022, No.11851302, No.11675243 and No.11761141011, and also supported by the Strategic Priority Research Program of the Chinese Academy of Sciences under grant No.XDB21010200 and No.XDB23000000. M.S. is supported by a grant from the General Research Fund by the Research Grants Council of Hong Kong SAR (Ref. 17305517).

\end{document}